%% file: arxiv_pmm.tex
\PassOptionsToPackage{unicode}{hyperref}
\PassOptionsToPackage{hyphens}{url}
\PassOptionsToPackage{dvipsnames,svgnames,x11names}{xcolor}
\documentclass[
  12pt]{article}

\usepackage{amsmath,amssymb}
\usepackage{iftex}
\ifPDFTeX
  \usepackage[T1]{fontenc}
  \usepackage[utf8]{inputenc}
  \usepackage{textcomp} 
\else 
  \usepackage{unicode-math}
  \defaultfontfeatures{Scale=MatchLowercase}
  \defaultfontfeatures[\rmfamily]{Ligatures=TeX,Scale=1}
\fi
\usepackage{lmodern}
\ifPDFTeX\else  
\fi
\IfFileExists{upquote.sty}{\usepackage{upquote}}{}
\IfFileExists{microtype.sty}{
  \usepackage[]{microtype}
  \UseMicrotypeSet[protrusion]{basicmath} 
}{}
\makeatletter
\@ifundefined{KOMAClassName}{
  \IfFileExists{parskip.sty}{%
    \usepackage{parskip}
  }{
    \setlength{\parindent}{0pt}
    \setlength{\parskip}{6pt plus 2pt minus 1pt}}
}{
  \KOMAoptions{parskip=half}}
\makeatother
\usepackage{xcolor}
\makeatletter
\ifx\paragraph\undefined\else
  \let\oldparagraph\paragraph
  \renewcommand{\paragraph}{
    \@ifstar
      \xxxParagraphStar
      \xxxParagraphNoStar
  }
  \newcommand{\xxxParagraphStar}[1]{\oldparagraph*{#1}\mbox{}}
  \newcommand{\xxxParagraphNoStar}[1]{\oldparagraph{#1}\mbox{}}
\fi
\ifx\subparagraph\undefined\else
  \let\oldsubparagraph\subparagraph
  \renewcommand{\subparagraph}{
    \@ifstar
      \xxxSubParagraphStar
      \xxxSubParagraphNoStar
  }
  \newcommand{\xxxSubParagraphStar}[1]{\oldsubparagraph*{#1}\mbox{}}
  \newcommand{\xxxSubParagraphNoStar}[1]{\oldsubparagraph{#1}\mbox{}}
\fi
\makeatother

\providecommand{\tightlist}{%
  \setlength{\itemsep}{0pt}\setlength{\parskip}{0pt}}\usepackage{longtable,booktabs,array}
\usepackage{calc} 
\usepackage{etoolbox}
\makeatletter
\patchcmd\longtable{\par}{\if@noskipsec\mbox{}\fi\par}{}{}
\makeatother
\IfFileExists{footnotehyper.sty}{\usepackage{footnotehyper}}{\usepackage{footnote}}
\makesavenoteenv{longtable}
\usepackage{graphicx}
\makeatletter
\def\maxwidth{\ifdim\Gin@nat@width>\linewidth\linewidth\else\Gin@nat@width\fi}
\def\maxheight{\ifdim\Gin@nat@height>\textheight\textheight\else\Gin@nat@height\fi}
\makeatother
\setkeys{Gin}{width=\maxwidth,height=\maxheight,keepaspectratio}
\makeatletter
\def\fps@figure{htbp}
\makeatother

\makeatletter
\@ifpackageloaded{caption}{}{\usepackage{caption}}
\AtBeginDocument{%
\ifdefined\contentsname
  \renewcommand*\contentsname{Table of contents}
\else
  \newcommand\contentsname{Table of contents}
\fi
\ifdefined\listfigurename
  \renewcommand*\listfigurename{List of Figures}
\else
  \newcommand\listfigurename{List of Figures}
\fi
\ifdefined\listtablename
  \renewcommand*\listtablename{List of Tables}
\else
  \newcommand\listtablename{List of Tables}
\fi
\ifdefined\figurename
  \renewcommand*\figurename{Figure}
\else
  \newcommand\figurename{Figure}
\fi
\ifdefined\tablename
  \renewcommand*\tablename{Table}
\else
  \newcommand\tablename{Table}
\fi
}
\@ifpackageloaded{float}{}{\usepackage{float}}
\floatstyle{ruled}
\@ifundefined{c@chapter}{\newfloat{codelisting}{h}{lop}}{\newfloat{codelisting}{h}{lop}[chapter]}
\floatname{codelisting}{Listing}

\makeatother
\makeatletter
\@ifpackageloaded{caption}{}{\usepackage{caption}}
\@ifpackageloaded{subcaption}{}{\usepackage{subcaption}}
\makeatother

\ifLuaTeX
  \usepackage{selnolig}  
\fi
\usepackage[]{natbib}
\usepackage{bookmark}

\IfFileExists{xurl.sty}{\usepackage{xurl}}{} 
\hypersetup{pdftitle={Title}, pdfauthor={Paul T. von Hippel}, pdfkeywords={3 to 6 keywords, that do not appear in the title},colorlinks=true,linkcolor={blue},filecolor={Maroon},citecolor={Blue},urlcolor={Blue},pdfcreator={LaTeX via pandoc}}

\newcommand{\anon}{1}

\begin{document}

\def\spacingset#1{\renewcommand{\baselinestretch}%
{#1}\small\normalsize} \spacingset{1}


\if1\anon { 
\title{\textbf{Imputing With Predictive Mean Matching Can Be Severely Biased
When Values Are Missing At Random}}
\author{Paul T. von Hippel\hspace{0.2cm}\\
 The University of Texas at Austin\\
}

\maketitle
} \fi

\if0\anon { \bigskip{}
\bigskip{}
\bigskip{}

\begin{center}
{\LARGE\textbf{Imputing With Predictive Mean Matching Can Be Severely
Biased When Values Are Missing At Random}}\textbf{ }
\par\end{center}

\medskip{}
} \fi

\bigskip{}

\begin{abstract}
Predictive mean matching (PMM) is a popular imputation strategy that imputes missing values by borrowing observed values from other cases with similar expectations. We show that, unlike other imputation strategies, PMM is not guaranteed to be consistent---and in fact can be severely biased---when values are missing at random (when the probability a value is missing depends only on values that are observed).

We demonstrate the bias in a simple situation where a complete variable $X$ is both  correlated with $Y$ and strongly predictive of whether $Y$ is missing. In simulation, we find that bias in the estimated regression slope can be as large as $-86\%$. As long as $X$ strongly predicts whether $Y$ is missing, bias persists even when we increase the sample size or reduce the correlation between $X$ and $Y$. To make the bias vanish, the sample must be large ($n$=1,000) \emph{and} $Y$ values must be missing independently of $X$ (i.e., missing completely at random).

Compared to other imputation methods, it seems that PMM requires larger samples and is more sensitive to the pattern of missing values. We cannot recommend PMM as a default approach to imputation. 
\end{abstract}
\vfill{}

\newpage\spacingset{1.8} 

\section{Predictive Mean Matching}\label{sec-intro}

 \emph{Multiple imputation} (MI) is a popular strategy for working with incomplete data. MI makes multiple copies of the incomplete data, fills in each copy's missing values with different plausible imputations, analyzes the multiple imputed datasets as though they were complete, and combines the results of the multiple analyses. 

MI estimates will be consistent if two conditions are met \citep{rubin1987multiple}: 
\begin{enumerate}
\tightlist
\item Values must be  \emph{missing at random} (MAR)---meaning that the probability a value is missing depends only on values that are observed. 
\item And the imputation model---the statistical model used to generate imputed values---must be correct. 
\end{enumerate}
In practice, however, imputation models are often only approximations. For example, a model that assumes a continuous normal distribution may be used to impute a variable that is discrete or skewed. Values imputed from such an incorrect model can be unrealistic in some cases. For example, we might impute negative values for a variable that is strictly positive, or fractional values for a variable that can only take integer values.

To make imputed values more realistic, many analysts use a technique called  \emph{predictive mean matching} (PMM) \citep{little1988missing,rubin1986statistical}. PMM fills in missing values by borrowing observed values from other
cases with similar expectations. For example, if a variable $Y$ has missing values, PMM uses linear regression to estimate the ``predicted mean'' $\hat{Y}$ of $Y$ given $X$. Then, for each case with missing $Y$, PMM imputes an observed $Y$ value sampled from a \emph{donor pool} of “matching” cases with similar values of $\hat{Y}$. 

In simple data where only $Y$ is missing and there is only one $X$ to use as a predictor, PMM reduces to imputing missing $Y$ values by sampling observed $Y$s from cases with similar values of $X$. That is the setting that we will focus on in this paper. However, PMM can use multiple predictors $X$, and PMM has been generalized to data where some $X$ variables are incomplete as well, and each variable is imputed conditionally on the others \citep{van2011mice}.

PMM has become widely available and popular. It is available in the core imputation commands for for SAS, SPSS, and Stata software. In R, the most popular imputation package,  \emph{mice}, has made PMM the default method for imputing continuous variables for at least 10 years \citep{van2015package} (from version 2.18 or earlier to version 3.18). 

No doubt PMM's popularity stems from the perception that values imputed using PMM look like observed values, because they are. If observed values are discrete, imputed values will also be discrete. If the observed variable is strictly positive, values imputed by PMM can never ben negative.

Despite its popularity, PMM's statistical properties are not well understood. No one has developed a theoretical argument explaining when and why PMM should produce consistent estimates. Although there have been several simulation studies of PMM, the results of these simulations have been mixed, and none has attempted to draw broad conclusions about the method's assumptions or basic soundness. While some simulation studies have reported that PMM ``generally worked well'' \citep{kleinke2017multiple} or ``is the only method that yields plausible imputations and preserves the original data distributions'' \citep{vink2014predictive}, some of the same studies, and others, have found that PMM can produce biased estimates under some simulated settings '\citep{vink2014predictive,landerman1997empirical}. Other simulations have focused on narrower questions, such as whether PMM was implemented correctly in software, or how many matching cases belong in the donor set from which imputed values are drawn \citep{morris2014tuning,allison2015predictive}.

In this article, we present evidence that PMM has a fundamental flaw. Unlike other imputation methods, PMM does not assure consistent estimates--and in fact, can produce severely biased estimates---when values are MAR. In our simulations, the only data where PMM produces unbiased estimates are large samples (n=1,000) where values are missing completely at random (MCAR), meaning that the probability a value is missing does not depend on any values, observed or unobserved. The MCAR assumption is rarely met, and when it is met, many other methods produce consistent estimates as well.

We illustrate the bias of PMM with a few scatterplots, then conduct a broad simulation to check the breadth of our conclusions. Because PMM can be severely biased when
data are MAR, we conclude that PMM is not a good default choice for handling missing data. Investigators considering PMM need to be aware of its liabilities.

\section{Example of bias from predictive mean matching}\label{section:graphical-example}

A fundamental assumption of PMM is that for every missing value, the data contain a set of observed values that makes plausible imputations. But this is not always the case. There are data where $Y$ values are MAR and yet, for some missing $Y$ values, there are no observed $Y$ values whose values of $\hat{Y}$ are close enough to be make plausible imputations. 

To construct such a data set, we start with a complete variable $X$ that is both strongly correlated with $Y$ and strongly predictive of whether $Y$ is missing. 

As an example, consider students in Mississippi, who must repeat third grade unless their score $X$ on a spring reading test exceeds a threshold  for passing. About 84\% of third graders pass the test on the first try. The rest take a retest---either an entirely different test or an alternate form of the initial test. So the retest is only observed for students in the bottom 16\% or so of the distribution of the initial score $X$; for the other 84\% of students, the retest score $Y$ is missing.

Note that the retest scores $Y$ are MAR. The probability that a retest score is missing depends only on the initial score $X$, which is fully observed.

What would happen if we tried to impute missing retest scores $Y$ conditionally on initial scores $X$? Individual students' scores are not publicly available, but we have a pretty good idea what they would look like. Test scores often have something close to a normal distribution, and the correlation between test and retest scores is typically near 0.8 \citep{von2019test}.

Accordingly, we simulated the scores of $n=200$ children by drawing from a standard bivariate normal distribution with a correlation of 0.8 between the test $X$ and retest $Y$. We then deleted the retest scores $Y$ of students scoring above the 16th percentile of the $X$ distribution, because these students would not take the retest $Y$.

\begin{figure}[htbp]
\centering \includegraphics[width=6.25in]{"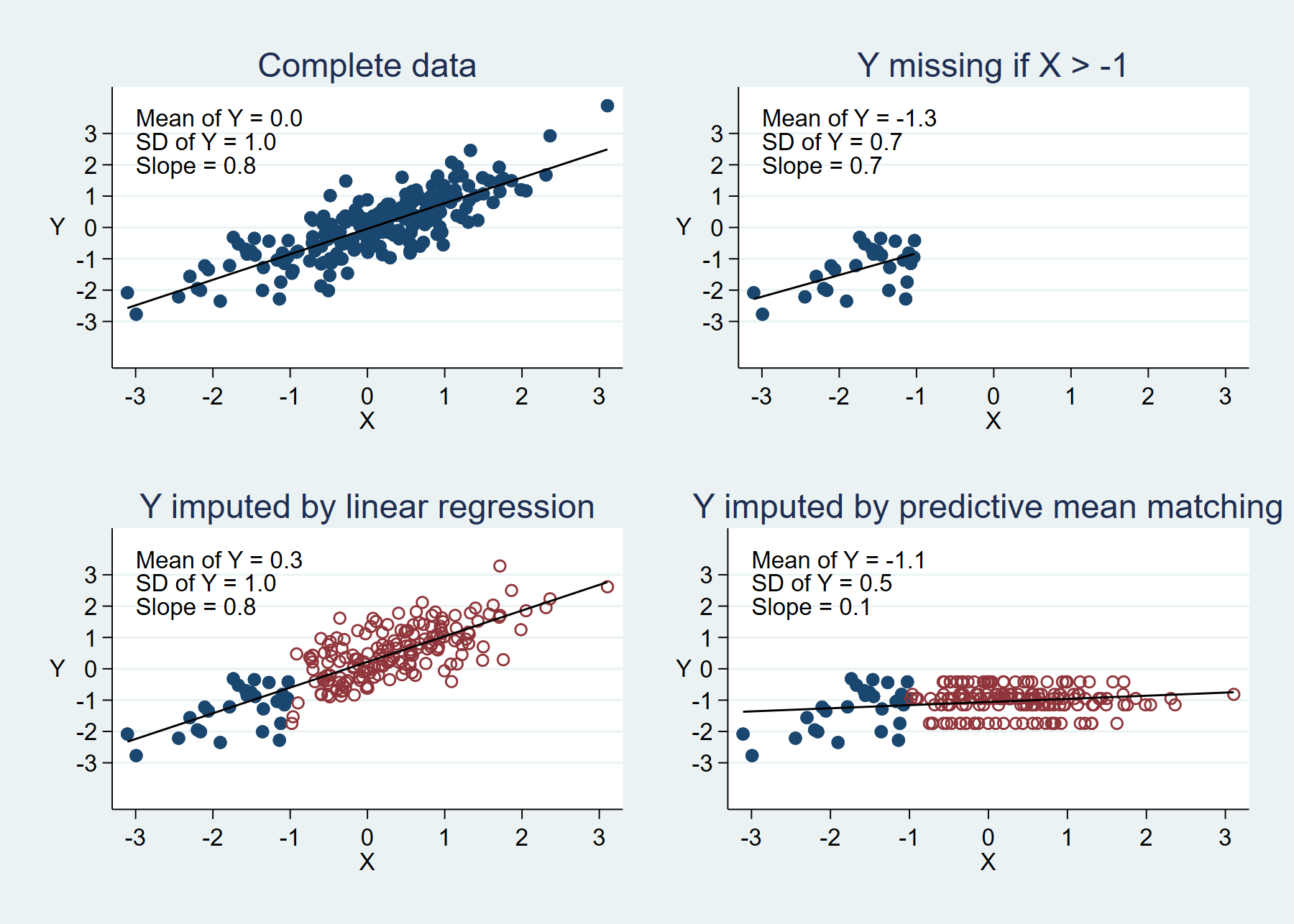"}
\caption{Simulated bivariate standard normal data with a 0.8 correlation between $X$ and $Y$, and $Y$ values missing if $X>-1$. Linear regression imputes the missing values well; predictive mean matching imputes them poorly.}
\label{fig:MAR-.8} 
\end{figure}

The upper half of Figure~\ref{fig:MAR-.8} illustrates the data: 
\begin{itemize}
\item The upper left panel shows complete test score data. The mean of the $Y$ values is 0, the SD 1, and the slope of $Y$ on $X$ is 0.8. (Since the variables are standardized, the slope is just the correlation when the data are complete.) 
\item The upper right panel deletes every $Y$ value if $X>-1$---that is, it deletes the retest scores of students who would not have taken the retest because their initial scores were in the top 84\% of the distribution. 
\end{itemize}

The lower half of Figure~\ref{fig:MAR-.8} illustrates two ways to impute the data: 
\begin{itemize}
\item The lower left imputes missing $Y$s from a linear regression with normal residuals. In this simple setting, with $X$ complete and $Y$ MAR, unbiased estimates of the regression parameters can be estimated by applying ordinary least squares to the cases with $Y$ observed \citep{little1992regression}. 
\item The lower right imputes missing $Y$s using PMM---imputing each missing $Y$ value by sampling from the 5 cases with the nearest value of $X$. Again, since there is only one regressor $X$, selecting from the cases with the closest $X$ values is equivalent to the more general recipe of selecting from the cases with the closest values of $\hat{Y}$.
\end{itemize}
Normal linear regression did an excellent job of imputing missing $Y$s. The imputed data forms an elliptical point cloud that looks very much like the point cloud of the complete data before $Y$ values were deleted. Estimates obtained from the imputed data are excellent as well. The $Y$s in the imputed data have an SD of 1.0 and a slope of 0.8---identical (to one decimal point) to the SD, slope, and correlation in the complete data. The mean of $Y$ in the imputed data is -0.3---just a bit lower than the mean of the complete $Y$ values before deletion.

Predictive mean matching, by contrast, butchered the data. The distribution of PMM-imputed scores looks nothing like the distribution of the complete data before $Y$ values were deleted. Instead of an elliptical cloud, the PMM-imputed values leave 5 horizontal stripes across the right side of the plot---implying that, above the 16th percentile of $X$, $X$ has no correlation with $Y$.

The slope in the PMM-imputed data is just 0.1, while the slope in the complete data was 0.8. That is, regression using PMM-imputed data underestimates the slope by 87\%---our simulation will show that this amount of bias is typical for data where $Y$ is missing when $X>-1$ Note that 87\% is also approximately the percentage of $Y$ values that were imputed (84\%). This is not coincidental. The slope through all the $Y$ values appears to be a weighted average of the slope through the observed $Y$ values (0.8) and the slope through the imputed $Y$ values (0), with weights approximately proportionate to the number of observed and imputed values. 

Other statistics are also negatively biased. The mean of $Y$ in the PMM-imputed data is 1.1 complete-data SDs below the complete-data mean of 0. And the SD of $Y$ in the imputed data is half the complete-data SD of 1. 

Because the SD is underestimated, the standard error of the mean is likely to be underestimated as well, because the SD is used in calculating the standard error. This will reduce the width and coverage of confidence intervals, as our simulation will show.

Why did predictive mean matching fail? It failed because in these data the observed $Y$ values were not a plausible pool from which to impute missing $Y$ scores. Borrowing retest scores $Y$ from 5 children who failed the initial test $X$ is not a valid way to impute the missing retest scores of readers who passed the initial test. Although the approach might work all right for student near the threshold for passing, the higher we get in the $X$ distribution, the less plausible the available observed $Y$ values become.

Note that PMM produced severely biased estimates  \emph{even though the data were MAR}. Theoretically, MAR guarantees consistent point estimates if the imputation model is correct---so the imputation model implied by PMM must be wrong. It is simply not the case that the expected values of the missing $Y$ values are close to 5 of the observed $Y$ values, as PMM assumes. 

To get good imputed values, a correctly specified model sometimes needs to extrapolate beyond the range of the observed values. Extrapolation is why the linear regression model was successful here.

\subsection{Varying the example}\label{subsection:varying-the-example}

The single example that we just presented is enough, by itself, to prove that PMM can produce biased estimates when data are MAR. But varying the example can help clarify what is causing the bias and what is likely make it better or worse. In the next section, we will vary the example systematically in a simulation experiment, but first we will discuss and illustrate a few variations more informally.

\begin{itemize}
\item \emph{Sample size}. Simply by inspecting Figure~\ref{fig:MAR-.8}, we can tell that increasing the sample size would not materially reduce the bias. The problem in the MAR example is not that there aren't  \emph{enough} observed $Y$ values; the problem is that there are  \emph{no} observed $Y$ values for high values of $X$. Increasing the sample size would not change this. 
\item \emph{Size of donor pool}. Changing the number of observed values from which the imputed values are sampled would also not materially reduce the bias. In Figure~\ref{fig:MAR-.8}, we used 5 observed values---a common default in PMM software---and this left 5 horizontal stripes across the right side of the graph. The problem is that the stripes are horizontal, not that there are 5 of them. Sampling fewer observed values, or more, would not tilt the slope of the stripes upward. 
\item \emph{Fraction of missing information}. The fraction of missing information is quite high in Figure~\ref{fig:MAR-.8}. Approximately 84\% of $Y$ values are missing, and the observed values are concentrated at one end of the regression line, providing less information about the slope than they would if they were more evenly distributed. Clearly the bias would be smaller if more $Y$ values are observed---but \emph{all} missing-data methods improve when the fraction of missing information is low. We chose to keep the fraction of missing information high, so that biases would be obvious in simple scatterplots and the power to detect them in our simulation experiment would be high. 
\end{itemize}

We constructed our example so that the relationship $X$ and $Y$ would be strong in two ways:
\begin{enumerate}
\item $X$ is strongly correlated with $Y$; and
\item $X$ is strongly predictive of whether $Y$ is missing.
\end{enumerate}
The question arises: what would happen to the bias if we weakened either of these relationships?

\begin{figure}[htbp]
\centering \includegraphics[width=6.25in]{"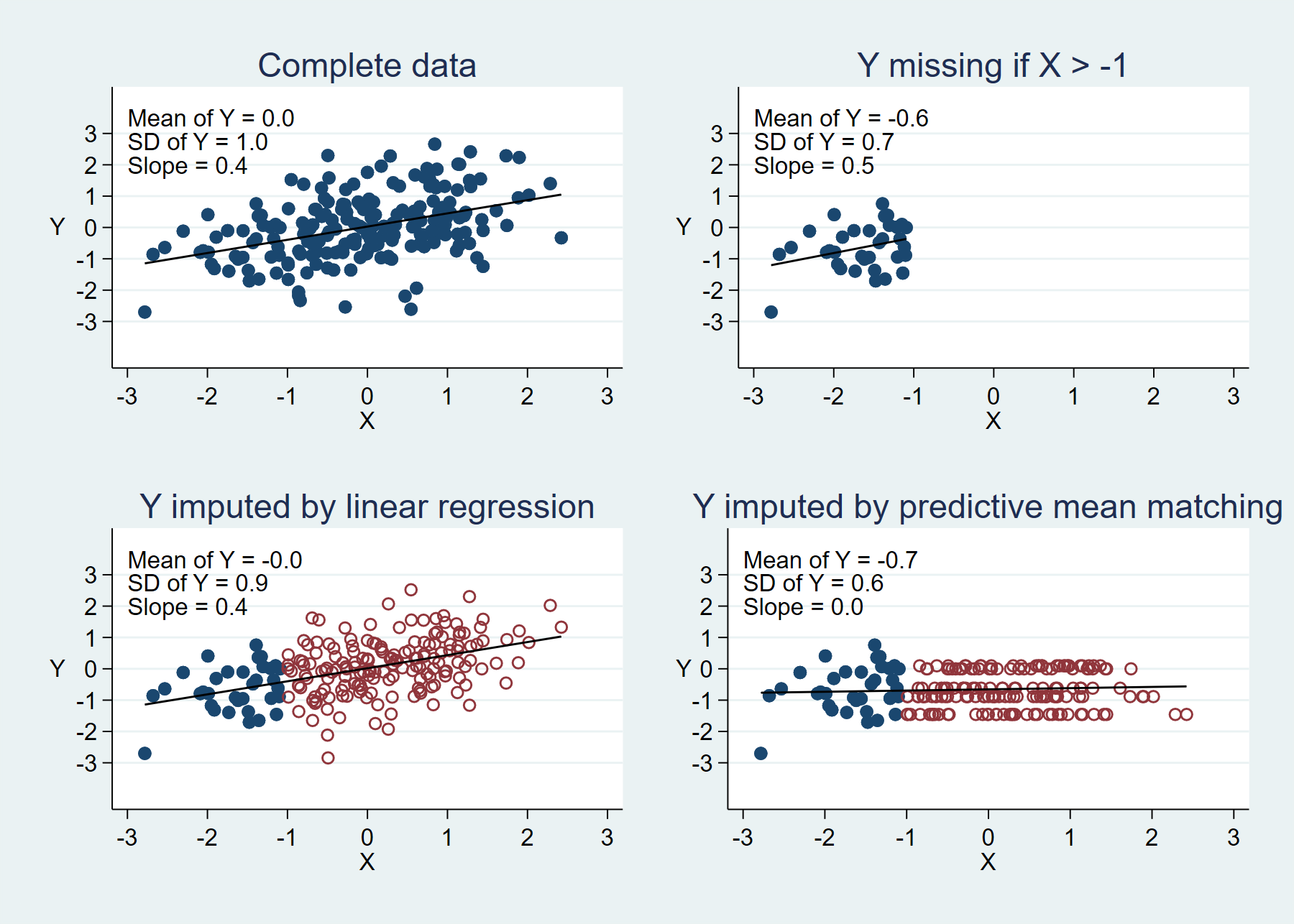"}
\caption{Simulated bivariate standard normal data with a 0.4 correlation between $X$ and $Y$ and $Y$ values missing whenver $X>1$. Relative to the true slope, the bias is just as severe as it was when the true slope was 0.8 in Figure \ref{fig:MAR-.8}}.
\label{fig:MAR-.4} 
\end{figure}

Figure \ref{fig:MAR-.4} illustrates what happens when we weaken the correlation from 0.8 to 0.4, while maintaining a MAR pattern where $Y$ is missing whenever $X>-1$. The bias of PMM remains severe. While the slope in the complete data is 0.4, and the slope in the regression-imputed data is 0.8 as well, the slope in the PMM-imputed data is only 0.1. 

In Figure \ref{fig:MAR-.4}, PMM underestimated the slope by 75\%, but on average, our simulation will show, PMM in MAR data underestimates a slope of 0.4 by about 86\%---the same fraction by which it underestimates a slope of 0.8. In other words, when we consider the bias relative to the true value of the slope, reducing the correlation does nothing to reduce the bias of the regression slope. The bias of the estimated mean and SD did improve, however, when the correlation weakened.

\begin{figure}[htbp]
\centering \includegraphics[width=6.25in]{"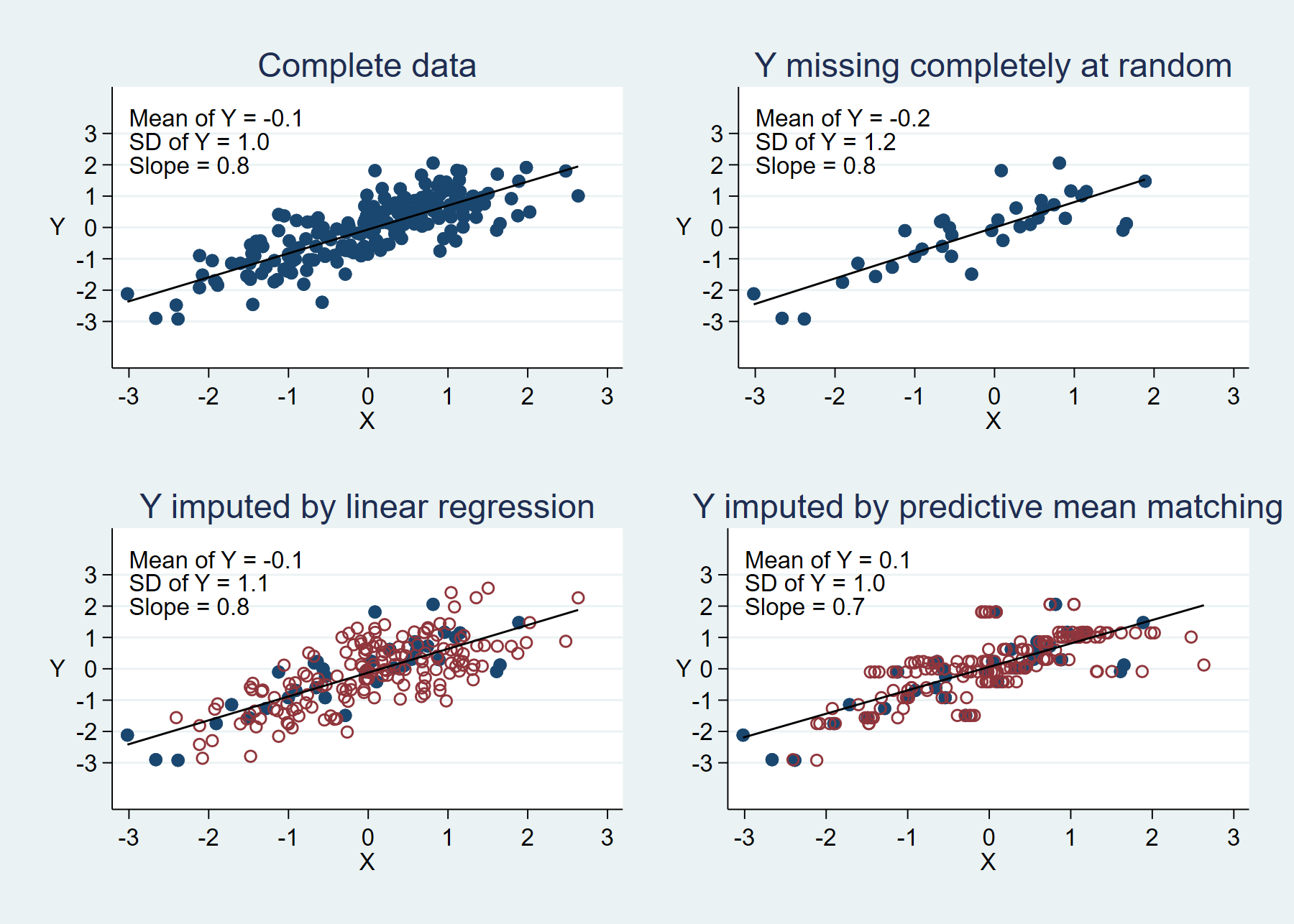"}
\caption{Simulated bivariate standard normal data with a 0.8 correlation between $X$ and $Y$ and 84\% of $Y$ values missing completely at random. PMM still biases the slope of the regression, but the bias is relatively mild and it fades as the sample size increases.}
\label{fig:MCAR-.8} 
\end{figure}

What happens when $Y$ values are missing completely at random (MCAR)---that is, when the probability that $Y$ is missing is a constant 16\%, regardless of the value of $X$? Figure~\ref{fig:MCAR-.8} shows the results for MCAR data with a correlation of 0.8. 

The slope is still slightly biased toward 0, because we are imputing points in horizontal strips that have a slope of 0 over short ranges. But the bias is much smaller---about $-15\%$ in MCAR data, compared to $-86\%$ in MAR data. Our simulation will show that this is typical. The simulation will also show that the bias fades as the sample size increases. That is, sample size matters for the bias of PMM in MCAR data, though sample size did not matter for bias in our MAR data.

\section{Simulation experiment}\label{section:simulation}

We conducted a simulation experiment to explore more systematically what properties of the data made the bias of PMM better or worse.

\subsection{Design}

We independently manipulated three properties of the data:
\begin{enumerate}
\item \emph{Correlation}. The correlation between $X$ and $Y$ could take values of 0.8, 0.4, or 0.
\item \emph{Missing pattern}. Values of $Y$ could be missing completely at random (MCAR) or only missing if $X<-1$ (MAR). Either way, $Y$ was missing for 84\% of observations on average. 
\item \emph{Sample size}. The number of observations $n$ could be 200 or 1,000.
\end{enumerate}
Independently manipulating these three properties gave us a total of $12=3\times2\times2$ experimental conditions. Within each condition, we simulated 500 incomplete datasets---enough to estimate the coverage of nominal 95\% confidence intervals to within a standard error of one percentage point.

We chose not to manipulate two properties discussed earlier---the size of the donor pool and the percentage of missing values that were missing---because manipulating these properties would increase runtime and clutter the results without providing much additional insight. 

\subsection{Multiple imputation}

Our scatterplots imputed the data just once, but in the
simulation experiment we used multiple imputation (MI), which imputes multiple copies of the incomplete data, then averages the point estimates across imputed copies \citep{rubin1987multiple}. 

Of course, averaging estimates across multiple imputations does not reduce any bias associated with PMM. If a single PMM-imputed dataset produces biased estimates, then averaging those estimates across multiple PMM-imputed datasets will be just as biased. 

We imputed each incomplete dataset $M=10$ times using the \emph{mi impute} command in Stata version 18.0. The command makes $M$ copies of the incomplete data and imputes each copy $m=1,...,M$ as follows:

\begin{itemize}
\item Draw estimated regression parameters $a_m, b_m, s_m^2$ at random from the posterior distribution of the parameters
\item For each missing value, estimate the predicted value $\hat{Y}==a_m + b_m X$. Then,
  \begin{itemize}
  \item if using regression imputation, impute $Y$ as $\hat{Y} + e_m$, where $e_m$ is a random draw from $N(0,s_m^2)$;
  \item if using PMM, impute $Y$ by sampling from the donor pool of the 5 observed $Y$ values with the closest values of $\hat{Y}$.
  \end{itemize}
\end{itemize}

After imputation, the \emph{mi estimate} command in Stata version 18.0 regressed $Y$ on $X$ in each imputed dataset, yielding $M$ point estimates of the slope $hat{beta}_m$ and other parameters, as well as an estimate $se_m$ that would estimate the slope's standard error consistently if the data were complete. It then combined these estimates to produce a single MI estimate of the slope, its standard error, and a 95\% confidence interval using standard formulas:

\begin{itemize}
\item The MI point estimate $\hat{\beta}_{MI}$ is just the average of the $M$ individual point estimates $hat{beta}_m$.
\item The MI standard error $se_{MI}$ is the square root of $V=W+(1+1/M)B$, where $W$ is the average of the squared standard errors $se_m$ and $B$ is the variance of the point estimates $hat{beta}_m$.
\item The MI confidence interval is $\hat{\beta}_{MI} \pm t \times se_{MI}$, where the degrees of freedom of the $t$ statistic is given by a small-sample formula that grows as the number of imputations $M$ grows or the fraction of missing information shrinks, but cannot exceed the degrees of freedom that the estimate would have if the data were complete \citep{barnard1999miscellanea}.
\end{itemize}

To save runtime, we imputed each incomplete dataset just $M=10$ times. Using more imputations can shorten confidence intervals and reduce random imputation error in individual MI estimates---but increasing the number of imputations would not reduce the bias or materially increase the coverage of MI confidence intervals \citep{von2020many}.

\subsection{Outcomes}

Our results highlighted two properties of the estimates:
\begin{itemize}
\item \emph{Average value of the point estimates.} This can be compared to the true value of the parameter.
\item \emph{\% relative bias of point estimates}. The difference between the average value of the point estimate and the true parameter value,
expressed as a percentage of the true parameter value.
\item \emph{\% coverage of confidence intervals}. The percentage of confidence intervals that actually contain the true parameter value. This can be compared to the nominal coverage rate of 95\%.
\end{itemize}
We report bias and coverage only for the regression slope. As our examples showed, if the regression slope is negatively biased, then the mean and standard deviation of $Y$ will be negatively biased as well.

\subsection{Results}

Table~\ref{tab:simulation_summary} summarizes the results of our simulation.

\begin{table}[hp]
\centering
\input{"Table_1.tex"}
\caption{Simulation comparing the bias of point estimates and coverage of nominal 95\% confidence intervals, for regression and PMM imputation in MAR and MCAR data with $X$ and $Y$ correlated at 0.8, 0.4, or 0. (\emph{Note.} When the true slope is 0, relative bias is not shown because it is not defined.)}
\label{tab:simulation_summary}
\end{table}

Across all experimental conditions, regression imputation produced valid estimates of the regression slope. Point estimates were unbiased, and confidence intervals covered the true regression slope in 94\% to 97\% of samples---close to the nominal level of 95\%. This is not surprising. Linear regression is the correct model for these data, and it can be shown analytically that regression imputation produces unbiased estimates of the regression slope when the data are MAR \citep{von2016new}.

The performance of PMM was worse than regression across every experimental condition. And of course PMM performed worse when values were MAR than when they were MCAR. 

\subsection{Performance of PMM when values were MAR}

When $Y$ was missing at random (MAR), predictive mean matching produced severely biased estimates of the regression slope. Whether the true slope was 0.8 or 0.4, the point estimates underestimated it by 85 to 86\% on average---a fraction similar to the fraction of $Y$ values that were missing. When the slope was 0, on the other hand, PMM estimated it without bias---but that was because the bias of PMM was toward zero. Like a broken clock that is right twice a day, an estimator that is biased toward zero will produced unbiased estimates only when zero is the parameter being estimated.

The coverage of PMM confidence intervals was also poor when $Y$ was MAR. When the true slope was 0.4 or 0.8, PMM's confidence intervals covered the true slope in just 0 to 4\% of samples. Even when the true slope was 0 and PMM point estimates were unbiased, PMM confidence intervals covered the true slope in just 48\% to 83\% of samples---well below the nominal level of 95\%. This is because PMM underestimated the residual variance, which contributes to the width of confidence intervals.

 As we predicted, increasing the sample did nothing to improve PMM estimates when values were MAR. The bias was practically the same whether the sample size was $n=200$ or $n=1,000$. And the coverage rate of confidence intervals actually declined as the sample size grew.

\subsection{Performance of PMM when values were MCAR}

When values were MCAR, PMM performed considerably better, and most PMM estimates did improve as the sample size grew. Under no condition, however, did PMM perform as well regression imputation. 

When the true slope was 0.4 or 0.8, PMM underestimated it by about 15\% on average in small samples ($n=200$),and the bias shrank to just 2\% to 3\% in larger samples ($n=1,000$). When the true slope was 0, PMM point estimates were unbiased---again because the bias of PMM was toward zero. 

Even when point estimates were unbiased, though, PMM confidence intervals covered the true slope in substantially less than 95\% of samples. When the true slope was 0.8, PMM confidence intervals covered the true slope in just 50 to 59\% of samples, and when the true slope was 0 or 0.4, PMM confidence intervals covered the true slope in only 68\% to 70\% of sample. Coverage improved with sample size if the true slope was 0.8, but not if the true slope was 0 or 0.4.

\subsection{Why did PMM confidence intervals under-cover?}
The poor coverage of PMM confidence intervals deserves some explanations. There are two reasons for poor coverage:
\begin{enumerate}
\item MI confidence intervals only estimate variability in the point estimates. They do not compensate for bias. When the point estimate of a slope has substantial bias, as it did for PMM in many simulated conditions, the confidence interval will cover the true slope less often than the nominal level.
\item Even when the estimated slope is unbiased, PMM may underestimate the residual variance of the regression, which is an important component of the standard error and confidence interval. When the residual variance is underestimated, the confidence interval will under-cover the true regression slope. It is possible to construct data where PMM \emph{overestimates} the residual variance, but in our simulations, PMM always underestimated the residual variance.
\end{enumerate}

\section{Conclusion}\label{sec-conc} 

Our results lead to some important conclusions:
\begin{itemize}
\item Unlike a correctly specified imputation model, PMM can produce severely biased estimates. When values are MAR, substantial bias can persist no matter how large the sample is.
\item Even when items are MCAR, PMM estimates can have non-negligible bias in small samples, though the bias fades as the sample gets larger.
\item Even when the PMM estimate of a regression slope is unbiased, confidence intervals can have poor coverage in both large and small samples.
\end{itemize}

We conclude that PMM is not a good default choice for handling missing data. PMM is not necessarily consistent when values are MAR, and may require a large sample of MCAR data to produce approximately unbiased estimates.

The poor performance of PMM in our simulation is disappointing, but perhaps it should not surprise us. After all, other efforts to make imputed values more realistic can also produce biased estimates. For example, if a dummy variable is imputed from a normal model, rounding imputed values to 0 or 1 can produce biased estimates of proportions and regression coefficients \citep{horton2003potential,allison2005discrete}. Similarly, if a strictly positive variable is imputed from a normal model, rounding negative imputed values up to zero can produce biased estimates, and re-imputing values until a positive value occurs can be even more biased \citep{von2013skewed}. 

In considering methods that try to ``correct'' unrealistic values, researchers should be clear about their goals for imputation. Often the goal is not to impute realistic values, but to get realistic estimates when we analyze the imputed data. For example, when we conduct a regression, all that matters is that the imputed variables have approximately the right means, variances, and covariances. By trying to correct unrealistic individual values, we may inadvertently make the means, variances, and covariances less realistic \citep{von2013skewed}. 

In the case of PMM, it is also important to recognize that the most realistic imputed values do not necessarily match values that are observed. It is sometimes necessary, as we have seen, to extrapolate beyond the range of observed values to get good imputations. A parametric model, such as linear regression, can do this. PMM cannot.

In our simulation, we compared PMM to regression imputation, and we recognize that were were playing on regression's home court. PMM is a semi-parametric method, and it is not surprising that it does not perform as well as a parametric method when the parametric model is correct. However, the issue was not just that PMM performed poorly relative to regression imputation. PMM also performed poorly in an absolute sense---producing severely biased estimates even when the data were MAR and the sample size was large.

On occasion, though, the investigator might be interested in quantities that are sensitive to the shape of the distribution---such as the percentiles, the skewness, or the Gini coefficient. In those settings, the poor fit of a simple parametric model might produce substantial biases \citep{von2013skewed}, but there is no guarantee that PMM will make the estimates better. Simple adjustments like rounding the imputed values often make estimates worse even when the parametric imputation model is a poor fit, and the limitations of PMM do not disappear if the distribution is idiosyncratic. If data are MAR, PMM can be inconsistent no matter what the true distribution is. Even when a parametric model fits the data poorly, then, it is not easy to know whether PMM is a desirable alternative. 

\section{Disclosure statement}

\label{disclosure-statement}

The authors have no conflicts of interest.

\section{Data availability statement}\label{data-availability-statement}

Stata code written for the simulations in this article is available at https://osf.io/tvzgk/.

\bibliographystyle{agsm}
\bibliography{bibliography}

\end{document}

%% file: Table_1.tex
\begin{tabular}{l l l c c c c}
\hline
& & & \multicolumn{2}{c}{n = 200} & \multicolumn{2}{c}{n = 1,000} \\
\cline{4-5} \cline{6-7}
& & & \multicolumn{4}{c}{\textbf{Imputation method}} \\
\cline{4-7}
Missing pattern & True slope & &  Regression & PMM & Regression & PMM \\
\hline

MAR & 0.8 & Mean slope estimate & 0.81 & 0.12 & 0.80 & 0.13 \\
    &     & \% relative bias     & 1    & -85  & 0    & -84  \\
    &     & \% CI coverage       & 98   & 0    & 94   & 0    \\
    & 0.4 & Mean slope estimate  & 0.41 & 0.06 & 0.40 & 0.07 \\
    &     & \% relative bias     & 2    & -86  & 1    & -84  \\
    &     & \% CI coverage       & 97   & 4    & 97   & 1    \\
    & 0   & Mean slope estimate  & -0.01& -0.01& -0.01& 0.01 \\
    &     & \% CI coverage       & 97   & 83   & 96   & 48   \\
MCAR & 0.8 & Mean slope estimate & 0.81 & 0.68 & 0.80 & 0.77 \\
     &     & \% relative bias     & 1    & -14  & 0    & -3   \\
     &     & \% CI coverage       & 95   & 50   & 94   & 59   \\
     & 0.4 & Mean slope estimate  & 0.41 & 0.34 & 0.41 & 0.39 \\
     &     & \% relative bias     & 2    & -16  & 2    & -2   \\
     &     & \% CI coverage       & 96   & 69   & 95   & 70   \\
     & 0   & Mean slope estimate  & -0.02& -0.02& -0.00& -0.00 \\
     &     & \% CI coverage       & 94   & 70   & 96   & 68   \\
\hline
\end{tabular}